# A HIGH-INTENSITY H- LINAC AT CERN BASED ON LEP-2 CAVITIES

M. Vretenar for the SPL Study Group, CERN, Geneva, Switzerland

*Abstract*

In view of a possible evolution of the CERN accelerator complex towards higher proton intensities, a 2.2 GeV H- linac with 4 MW beam power has been designed, for use in connection with an accumulator and compressor ring as proton driver of a muon-based Neutrino Factory. The high-energy part of this linac can use most of the RF equipment (superconducting cavities and klystrons) from the LEP collider after its decommissioning at the end of 2000. Recent results concerning low-beta superconducting cavities are presented, and the main characteristics of the linac design are described. The complete linac-based proton driver facility is outlined, and the impact on the linac design of the requirements specific to a Neutrino Factory is underlined.

## 1. THE LEP-2 RF SYSTEM

The decommissioning of the CERN LEP $e^+ e^-$ collider at the end of 2000 will pave the way to the construction of the Large Hadron Collider (LHC), but will also present the unprecedented challenge of the removal, storage or disposal, and possible recycling of the huge amount of valuable LEP equipment.

A particularly valuable item is the 352.2 MHz superconducting RF system built for the Phase 2 of LEP, consisting of 288 four-cell cavities (Figure 1) operating at 4.5 °K and powered by 36 1.3 MW CW klystrons. It delivers a total accelerating voltage of about 3 GV to the electron beam. Eight more klystrons are used to power the normal-conducting RF system of LEP, for a total of 44 klystrons installed in the machine. Most of the superconducting cavities (272) were produced using the technique developed at CERN of sputtering a thin film of niobium onto copper [1]. The cavities were initially designed for a gradient of 6 MV/m, and during the 1999 run they achieved an average gradient of 7.5 MV/m, with up to 9 MV/m in some cavities [2]. In the basic LEP configuration, each klystron feeds 8 cavities via an array of magic tees, equipped with circulators and loads. Four cavities are grouped in a cryostat. The cavities and the cryostats are fully equipped with slow and fast tuners, power couplers matched for a beam current of 10 mA, high-order-mode couplers, superinsulation and insulation vacuum tanks.

The present plans foresee to store most of the RF material for possible future use. The cryogenic system of LEP will be used for the LHC magnets.

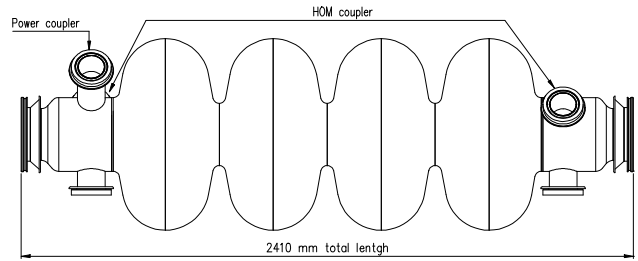

Figure 1: The LEP-2 accelerating cavity

## 2. A LINAC BASED ON LEP CAVITIES

Some proposals for re-using this expensive hardware have been made, such as for a Free Electron Laser [3] or to build the ELFE machine on the CERN site, a recirculating electron linac for nuclear physics [4].

An early proposal already opened the perspective of using the LEP cavities in a high beam power superconducting linac driving a hybrid reactor [5-7].

The main limitation for using these cavities in proton linacs comes from the fact that they are designed for $\beta=1$, their transit time factor drastically decreasing for a proton beam at low beta. Figure 2 shows a calculation of the effective cavity gradient as function of energy that can be reached by LEP cavities operating at a nominal gradient of 7.5 MV/m. While in principle they can be used for proton acceleration from about 500 MeV, they become efficient and economically justified only from about 1 GeV, i.e. in an energy range beyond the usual requirements of high-power linacs for spallation sources, transmutation or hybrid reactors.

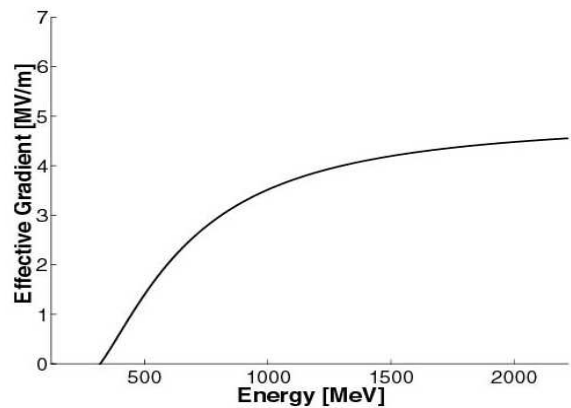

Figure 2: Effective gradient of the LEP-2 cavities as a function of energy

As will be seen in the following, applications of linacs for physics research at energy > 1 GeV exist, but they require beam powers of only a few MW and a well-defined time structure of the beam. This imposes a pulsed operation mode that has to be optimised to achieve a reasonable mains-to-RF efficiency. High duty cycles are preferable because they reduce the impact of the static cryogenic losses, and long pulses minimise the relative effect of the RF power lost during the relatively long (1-2 ms) pulse rise time, when all the power is reflected from the couplers.

However, the LEP cavities are well suited for pulsed operation because of the inherent rigidity of the copper cavity structure and of the relatively low gradient that make them less sensitive to Lorentz force detuning and vibration problems. The large (241 mm) aperture is particularly useful for machines sensitive to beam losses.

## 3 APPLICATIONS OF A 2 GEV LINAC

The first proposal to replace the present 50 MeV linac and the 1.4 GeV Booster in the CERN proton injector chain with a Superconducting Proton Linac (SPL) dates from 1996 [8]. In the original scheme this machine was intended to accelerate mainly protons, and although in the following studies the advantages of a common H$^-$ operation for all the users became clear, the title of SPL has been maintained. A first feasibility study [9] considered a 2 GeV H$^-$ SPL equipped with LEP cavities from 1 GeV energy, injecting at 0.8 Hz repetition rate into the Proton Synchrotron (PS) ring. This new injector would have several benefits over the present injection scheme for the LHC:
- a factor 3 increase in the brightness of the proton beam (density in transverse phase space) delivered by the PS injector complex, due to the lower space charge tune shift at injection, which is an advantage for LHC;
- the potential for improving the peak intensity in the PS for experiments requiring a high proton flux;
- the reduction of injection losses with charge exchange injection and a chopped linac beam;
- the replacement of the PS injectors by modern and standard equipment.

However, such a machine would be fully justified only when pulsed at a higher rate, and the original feasibility study aimed somehow arbitrarily for 5% duty cycle.

On the basis of this preliminary study, some user communities have shown their interest in a high-intensity facility at CERN. Firstly there is the strong demand for second generation radioactive nuclear beam facilities in Europe. The SPL could easily become the driver of a facility based at CERN that would profit from the experience gained at ISOLDE. The mean current required is about 100 µA, preferably distributed in many low intensity pulses and at a variable energy.

Secondly, strong interest has been recently shown by the physics community for the high-intensity high-quality neutrino beams that can be provided by a Neutrino Factory based on a muon decay ring. CERN has recently started a study on the technological challenges of such a Neutrino Factory, that resulted in the CERN Reference Scenario of Figure 3 [10].

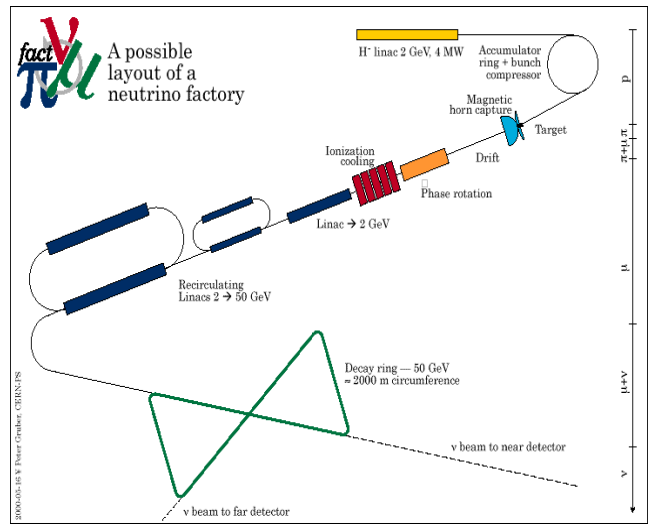

Figure 3: Possible layout of a Neutrino Factory

The main challenges for this machine come from the need for high neutrino fluxes. The interest for physics starts from some $10^{21}$ neutrinos/year, that can be obtained with about 4 MW beam power from a driver accelerator delivering protons on a target, producing pions which decay into muons. After cooling and acceleration, the muons are stored in a decay ring where they generate two intense neutrino beams.

Simulations of particle production in the target indicate that the number of pions is approximately proportional to beam power for energies $\geq$ 2 GeV. This suggests that a low-energy linac-based driver constitutes a viable alternative to conventional high-energy, fast-cycling synchrotrons. The HARP experiment at CERN [11] is intended to provide experimental data on pion production at different energies and from different targets, for a final confirmation of the low energy choice.

Muon collection, cooling system and decay ring impose a well-defined time structure for the beam on target. This requires two rings after the linac, an Accumulator to produce a 3.3 µs burst of 140 bunches at 44 MHz (the frequency of the muon phase rotation section) and a Compressor to reduce the bunch length to 3 ns [12]. The rings have been designed to fit in the existing ISR tunnel. Space charge and beam stability are their major design concerns. To reduce space charge tune shift at injection into the accumulator, the linac bunch length has to be stretched in the transfer line, from about 30 ps to 0.5 ns, by means of two bunch rotating cavities.

# 4 THE SPL H⁻ LINAC DESIGN

## 4.1 Main Parameters

The parameters of this machine (Table 1) had to take into account the optimum operating conditions of the superconducting cavities discussed in Section 2, and are mainly determined by the needs of the Neutrino Factory, by far the most demanding user in terms of particle flux and time structure of the beam pulses.

Table 1 Main linac design parameters

| Particles | H⁻ | |
|---|---|---|
| Kinetic Energy | 2.2 | GeV |
| Mean current during pulse | 11 | mA |
| Repetition frequency | 75 | Hz |
| Beam pulse duration | 2.2 | ms |
| Number of particles per pulse | $1.51 \times 10^{14}$ | |
| Duty cycle | 16.5 | % |
| Mean beam power | 4 | MW |
| RF Frequency | 352.2 | MHz |
| Chopping factor | 42 | % |
| Mean bunch current | 18 | mA |
| Transv. emittance (rms, norm.) | 0.6 | μm |

The mean current during the pulse of 11 mA has been selected as a compromise between the number of klystrons needed in the superconducting section, the efficiency of the feedback loops, the number of turns injected into the accumulator and the power efficiency of the room temperature section. The input couplers of the LEP cavities are already matched for this current.

The linac energy of 2.2 GeV, the repetition rate of 75 Hz and the corresponding pulse length of 2.2 ms are a compromise between the optimum operating conditions of the superconducting cavities and the need to limit the number of turns injected into the accumulator, 660 in the present scenario.

A chopper in the low energy section is used to minimise losses at injection in the accumulator and at the transfer between the rings. In the present design, 42% of the beam is taken out at the chopper position, leading to a source current and a bunch current of 18 mA. This value is within reach of present H⁻ sources and is well below the limits of space-charge dominated beam dynamics.

The layout of the linac is shown in Figure 4 and key data are given in Table 2.

Table 2 Layout data of the SPL H⁻ linac

| Section | Output Energy (MeV) | RF power (MW) | Nb. of klystrons | Nb. of tetrodes | Length (m) |
|---|---|---|---|---|---|
| Source | 0.045 | - | - | - | 3 |
| RFQ1 | 2 | 0.25 | 1 | - | 2 |
| Chopper | 2 | - | - | - | 3 |
| RFQ2 | 7 | 0.6 | 1 | - | 5 |
| DTL | 120 | 8.7 | 11 | - | 78 |
| SC-lowβ | 1080 | 10.6 | 12 | 74 | 334 |
| SC–LEP | 2200 | 12.3 | 18 | - | 357 |
| Total | | 32.5 | 43 | 74 | 782 |

The LEP RF frequency of 352 MHz has been maintained for the whole linac. The choice of this frequency for the low-beta superconducting cavities that have to be built for the SPL allows to apply the sputtering fabrication technique to the cavity and to use couplers and cut-off tubes recuperated from LEP units.

This frequency provides the additional flexibility that klystron or tetrode amplifiers can be used in the RF system. Klystrons can feed the high-power cavities (room temperature and high-beta superconducting), with a limited number of cavities connected to the same klystron. Individual 65 kW tetrode amplifiers can, instead, feed the low-beta superconducting cavities, thus avoiding the potential dangers at low beam energy of a field stabilisation based on the vector sum of many cavity signals.

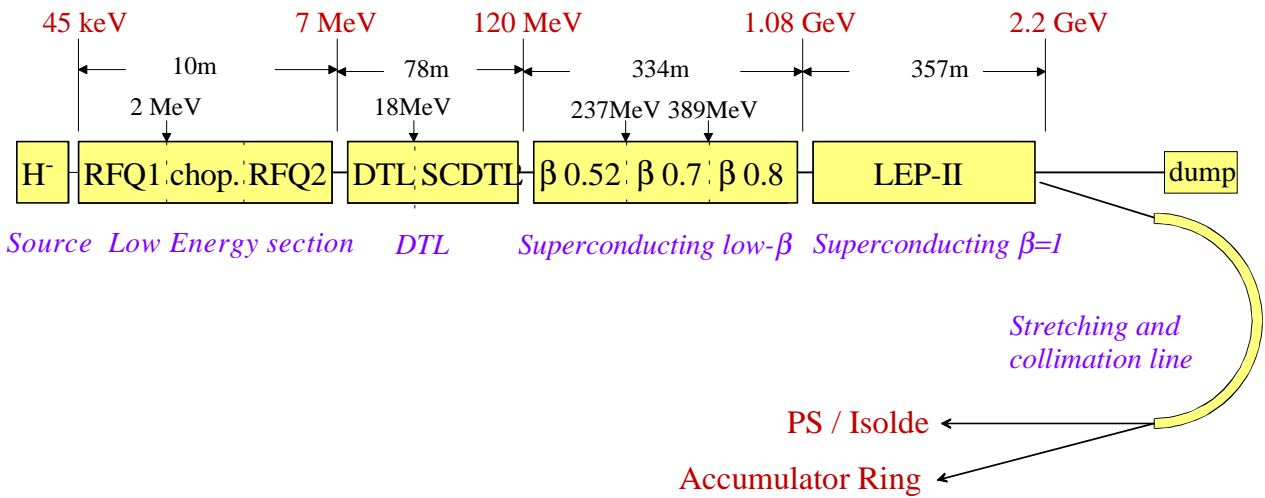

Figure 4 : Layout of the SPL H⁻ linac

Keeping beam losses below the limit for hands-on maintenance (1 W/m) has been a design issue from the beginning. The main principles were a design of the linac optics without excessive jumps in the focusing parameters, to avoid the formation of halo from mismatch at the transitions, a particular care to avoid crossing resonances, and finally, the preference for large apertures in spite of some reductions in shunt impedance. Wherever possible, losses will be concentrated on localised dumps by means of collimators.

### 4.2 Room Temperature Section

The design source current, 18 mA, is well within the range of existing H$^-$ sources, while the required pulse length and duty cycle are more challenging as compared to existing sources. Reliability is also an important concern. The study of an H$^-$ ECR source that could meet the SPL parameters has been started, and collaborations are envisaged.

The fast chopper (2 ns rise time) at 2 MeV could be a travelling-wave stripline structure similar to the LANL design [13]. An analysis of the options for the 1 kV pulse amplifier indicates that a combination of vacuum tubes driven by fast Mosfets could provide the required rise and fall times [14]; this will be tested on a prototype.

The DTL starts at 7 MeV, and is composed of two standard Alvarez tanks up to an energy of 18 MeV, followed by a section of Side-Coupled DTL (SCDTL) [15], 2-gap tanks connected by off-axis coupling cavities, with quadrupoles placed between tanks. The 352 MHz structure going up to 120 MeV is made of 98 small tanks grouped in 9 chains, each one powered by a klystron [16].

### 4.3 Superconducting Section

The superconducting part of the linac is composed of four sections made of cavities designed for β of 0.52, 0.7, 0.8 and 1 respectively. The LEP-2 cavities are used at energies above 1 GeV. The cavities at β=0.52 and β=0.7 contain 4 cells, whilst the beta 0.8 cavities are made of 5 cells, to re-use the existing LEP cryostats.

The main parameters of the superconducting section are summarised in Table 3. It has been assumed that the LEP cavities will operate at 7.5 MV/m, while for the newly-built β=0.8 cavities, cleaning procedures to achieve high gradients can be applied and a design gradient of 9 MV/m can be foreseen. During tests, a β=0.8 cavity has already reached gradients of 10 MV/m [17]. The transition energies between sections are defined in order to have the maximum effective accelerating gradient and to minimise phase slippage.

Table 3 Superconducting linac section

| Beta | $W_{out}$ (MeV) | Gradient (MV/m) | Cavities | Cryost. | Length (m) |
|------|-----------------|-----------------|----------|---------|------------|
| 0.52 | 237 | 3.5 | 42 | 14 | 101 |
| 0.7 | 389 | 5 | 32 | 8 | 80 |
| 0.8 | 1080 | 9 | 48 | 12 | 153 |
| 1 | 2200 | 7.5 | 108 | 27 | 357 |

The cavities at β=0.7 and β=0.8 can be built of niobium sputtered on copper. This technique, developed at CERN, has many advantages with respect to bulk Niobium for large productions:
a) the cost of the raw material is much lower, giving the possibility to go for low frequencies where the iris aperture is large, relaxing the mechanical tolerances and reducing the probability of beam losses;
b) Nb/Cu cavities can be operated at 4.5 °K with Q-factors of more than $10^9$, simplifying the design of the cryostats and of the power coupler;
c) the excellent mechanical properties of copper ensure a better thermal and mechanical stability.

A development programme was started at CERN in 1996 to investigate the feasibility of the production of cavities in the β range 0.5–0.8. The main results are the prototypes of a 5-cell β=0.8 cavity and of a 4-cell β=0.7 cavity (shown in Fig. 5), that have achieved satisfactory Q-values at high gradient (Fig. 6) [18,19]. Attempts to sputter cavities at β<0.7 were not successful, thus the 42 cavities at β=0.52 have to be made of bulk niobium.

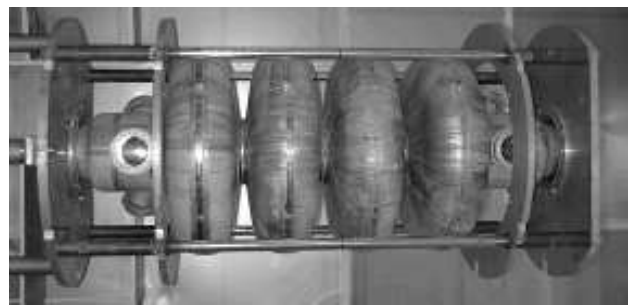

Figure 5: The prototype 4-cell β=0.7 cavity

Particular attention has been given to the pulsed operation of the superconducting cavities. Feedback loops are foreseen to minimise the effect on the beam of cavity vibrations and of Lorentz forces. In the β=0.8 and β=1 sections, where one klystron feeds 4 and 6 cavities respectively, the compensation has to be made on the vector sum. Simulations show that random oscillations of the cavity frequency of up to 40 Hz amplitude can be tolerated, without increasing the energy spread of the beam outside the ± 10 MeV corresponding to the acceptance of the accumulator [20].

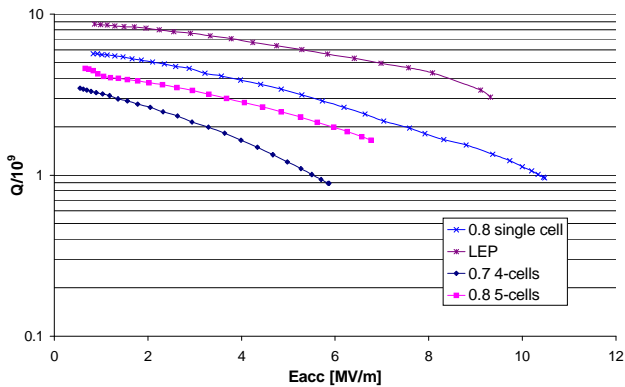

Figure 6: Q vs. gradient of the sputtered-Nb cavities

Multi-particle simulations of the beam dynamics in the superconducting section show a stable behaviour in the presence of errors and mismatch of the input beam [21].

### 4.4 Layout on the CERN site

After considering some possible locations for the SPL around the CERN Meyrin site, the option shown in Fig. 7 has been retained. Placing the linac and the parallel klystron gallery in an area immediately outside of the CERN fence on the Swiss side offers the advantages of an economic trench excavation, of minimum impact on the environment (the site is presently an empty field), of a simple connection to the ISR tunnel and to the PS through existing tunnels, and of an easy access from the road along the fence.

The infrastructure for electricity, water cooling and cryogenics makes a maximum use of existing facilities on the Meyrin site.

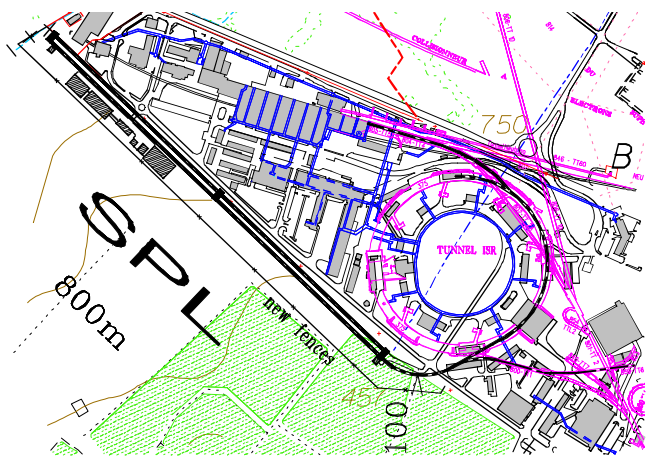

Figure 7: Layout of the SPL on the CERN site

## 5 CONCLUSIONS

About 40% of the LEP-2 cavities, 57% of the cryostats and all the klystrons plus other RF and HV equipment can be used to construct a 2.2 GeV $H^-$ linear accelerator on the CERN site that would improve the beam brightness and intensity of the PS ring, provide a flexible and powerful beam source for a second generation radioactive beam facility and constitute the first step towards a powerful Neutrino Factory.